\def\+{{(+)}}  \def\-{ {(-)} }   \def\0{ {(0)} }
\def\1{ {(1)} }  \def\2{ {(2)} }
           
\documentstyle[12pt]{article}

\def\be{\begin{equation}}             \def\ee{\end{equation}}
\def\ba{\begin{array}{rcl}}           \def\ea{\end{array}}
\def\beqa{\begin{eqnarray} }          \def\eeqa{\end{eqnarray} }
\def\beqalign{\begin{eqalign}}        \def\eeqalign{\end{eqalign}}

\def\bsubeq{\begin{subequations}}     \def\esubeq{\end{subequations}}
\def\bitem{\begin{itemize}}           \def\eitem{\end{itemize}}
\begin{document}

\title{\bf Non-Archimedean Geometry and\\ Physics on Adelic  Spaces
\thanks{Talk presented at the  Workshop {\it Contemporary Geometry
and Related Topics}, Belgrade, May 15-21, 2002.  } }

\author{Branko DRAGOVICH\thanks{e-mail address: dragovich@phy.bg.ac.yu}\ \ \\
     {\it Institute of Physics}\\ {\it P.O.Box 57, 11001 Belgrade,
      Serbia and Montenegro}}

\date{}
\maketitle
\begin{abstract}
This is a brief review article of various  applications of
non-Archimedean geometry, $p$-adic numbers and adeles in modern
mathematical physics.
\end{abstract}

\section{Introduction} 

It is well known that theoretical physics is strongly related to
mathematics. Space, time and matter  are basic concepts in all
physical theories. They have become usually profound and gradually
unified in new theories using more general mathematical tools. For
example, transition from nonrelativistic to relativistic
kinematics required to pass from Euclidean space and time to the
Minkowski space. To describe phenomena in strong gravitational
fields and accelerated frames, general theory of relativity was
discovered, where space-time is described by pseudo-Riemannian
geometry which is related to the distribution of matter. Dynamics
in quantum mechanics can be regarded as motion of a particle in a
phase space $(x,k)$  with symplectic geometry and the Heisenberg
uncertainty $\Delta x^i \Delta k^j \geq \frac{\hbar}{2}
\delta^{ij}$, where $\hbar =\frac{h}{2\pi}$ is the reduced Planck
constant. In recent years,  noncommutative geometry based on
relation $[x^i,x^j] \neq 0$ has attracted a significant interest
in quantum theory.

According to various considerations, which take together quantum
and gravitational principles, there is a restriction on empirical
accuracy  of physical variables due to the relation
$$
\Delta x \geq \ell_0 = \sqrt {\frac{\hbar G}{c^3}} \sim 10^{-33}
cm,     \eqno(1.1)
$$
where $\Delta x$ is an uncertainty measuring a distance, $\ell_0$
is the Planck length, $G$ is Newton's gravitational constant and
$c$ is the speed of light in  vacuum. The uncertainty (1.1) means
that one cannot measure distances smaller than $\ell_0$. Since
this result is derived assuming that space-time consists of real
points and has an Archimedean geometry, it becomes desirable to
employ also non-Archimedean geometry based on $p$-adic numbers.
Quite natural framework to consider simultaneously real
(Archimedean) and $p$-adic (non-Archimedean) spaces is by means of
an adelic space.

In this paper,  at an introductory level I briefly review some
basic characteristics of non-Archimedean geometry, $p$-adic
numbers and adeles,, as well as their use in some parts of modern
mathematical physics.

\section{Non-Archimedean Geometry and $p$-Adic Numbers}

Recall that having two segments of straight line of different
lengths $a$ and   $b$, where $a<b$, one can overpass the longer
$b$ by applying the smaller $a$ some $n$-times along $b$. In other
words, if $a$ and $b$ are two positive real numbers and $a<b$ then
there exists an enough large natural number $n$ such that $na>b$.
This is an evident property of the Euclidean spaces (and the field
of real numbers), which is known as Archimedean postulate, and can
be extended to the standard Riemannian spaces. One of the axioms
of the metric spaces is the triangle inequality which reads:
$$
d(x,y) \leq d(x,z) + d(z,y),                  \eqno(2.1)
$$
where $d(x,y)$ is a distance  between points $x$ and $y$. However,
there is a subclass of metric spaces for which triangle inequality
is stronger in such way that:
$$
d(x,y) \leq max \{d(x,z), d(z,y)  \}   \leq d(x,z) + d(z,y).
\eqno(2.2)
$$
Metric spaces with strong triangle inequality (2.2) are called
non-Archimedean or ultrametric spaces.

Since a measurement means quantitative comparison of a given
observable with respect to a fixed value taken as its unit, it
follows that a realization of the Archimedean postulate is
practically equivalent to the measurements of distances. According
to the uncertainty (1.1), it is not possible to measure distances
shorter than $10^{-33} cm$ and consequently there is no place for
an Archimedean geometry beyond the Planck length. By this  way,
standard approach to quantum gravity, which is based only on
Archimedean geometry and real numbers, predicts its own breakdown
at the Planck scale. Hence, a new approach, which takes into
account not only Archimedean but also a non-Archimedean geometry,
seems to be quite necessary. The most natural ambient to realize
both of these geometries is an adelic space, which is a whole of
real and all $p$-adic spaces.

Any set with trivial metric ($ d(x,y) =1$ if $x \neq y$ and
$d(x,y) =0 $ if $x=y$) is a simple example of the non-Archimedean
space. Set of all real polynomials ${\bf R}[X]$, for which metric
is defined by a suitable valuation, is another example of the
non-Archimedean space. Namely, for a nonzero polynomial $f \in
{\bf R}[X]$  given by $f = a_n X^n +\cdots + a_1 X + a_0, \ \ a_n
\neq 0$, one can define the degree of  $f$ as $d(f) = n$, and
$d(f) = - \infty$ if $f$ is the zero polynomial. Then a norm can
be introduced as $|f| = \rho^{d(f)}$ if $f \neq 0, \ $  and $ |f|
=0$ if $f=0$, where $\rho$ is a real number greater than 1. Since
$|f+g|\leq max\{ |f|, |g| \}$ and $|fg|=|f||g|$, where $f, g \in
{\bf R}[X],$ this is a non-Archimedean valuation which gives a
non-Archimedean geometry by $d(f,g) = |f-g|.$ This can be extended
to the field ${\bf R}(X)$ of rational functions. On the field of
complex numbers ${\bf C}$ there exist infinitely many inequivalent
non-Archimedean  valuations which make ${\bf C}$ into complete
valued fields (se \cite{schikhof} p. 46) with non-Archimedean
geometries. Hyperreal numbers in nonstandard analysis also have
non-Archimedean properties. In the sequel we will restrict to the
spaces of $p$-adic numbers as presently the most important class
of  non-Archimedean geometries. In the rest of  this section, a
brief review of some basic properties of $p$-adic numbers and
their functions will be presented.

There are physical and mathematical reasons to introduce $p$-adic
numbers starting with the field of rational numbers ${\bf Q}$ and
performing  completions with respect to its non-Archimedean
valuations. From physical point of view, numerical results of all
experiments and observations are some rational numbers. From
algebraic point of view, ${\bf Q}$ is the simplest number field of
characteristic $0$. Recall that any $0\neq x\in {\bf Q}$ can be
presented as infinite expansions into the two essentially
different forms \cite{schikhof}:
$$
 x = \sum_{k=n}^{-\infty} a_k 10^k, \ \  a_k = 0,1,\cdots ,9,
\ \ a_n \neq 0,   \eqno(2.3)
$$
$$
  x =  \sum_{k=m}^{+\infty} b_k p^k, \ \  b_k = 0,1,\cdots ,p-1,
\ \  b_m \neq 0,  \eqno(2.4)
$$
where (2.3) is the ordinary one to the base $10$,  (2.4) is
another to the base $p$ ($p$ is any prime number), and  $n,\ m$
are some integers which depend on $x$. The above representations
(2.3) and  (2.4) exhibit the usual repetition of digits, however
the expansions are in the mutually opposite directions. The series
(2.3) and  (2.4) are convergent with respect to the metrics
induced by the usual absolute value $| \cdot |_\infty$ and
$p$-adic absolute value ($p$-adic norm, $p$-adic valuation) $|
\cdot |_p$, respectively. Due to the Ostrowski theorem these
valuations exhaust all possible inequivalent non-trivial norms on
${\bf Q}$. Performing completions to all non-trivial norms, {\it
i.e.} allowing all possible combinations for digits, one obtains
standard representation of real and $p$-adic numbers in the form
(2.3) and (2.4), respectively. Thus, the field of real numbers
${\bf R}$ and the fields of $p$-adic numbers ${\bf Q}_p$ exhaust
all number fields which may be obtained by completion of ${\bf
Q}$, and consequently  contain ${\bf Q}$ as a dense subfield.
Since $p$-adic norm of any term in (2.4) is $| b_k p^k |_p =
p^{-k}$ if $b_k \neq 0$, geometry of $p$-adic numbers is the
non-Archimedean one, {\it i.e.} strong triangle inequality $| x+y
|_p \leq max (| x |_p, | y |_p) $ holds and $| x |_p =p^{-m}$.
${\bf R}$ and ${\bf Q}_p$ have many distinct algebraic and
geometric properties.

Unlike the real case, there are different algebraic extensions for
all orders of $p$-adic algebraic equations. Algebraic closure of
${\bf Q}_p$ is an infinite ${\bf Q}_p$-vector space and it is not
metrically complete. After completion it becomes the field of
$p$-adic complex numbers ${\bf C}_p$, which is also algebraically
closed, and is a $p$-adic analogue of the ordinary ${\bf C}$. This
${\bf C}_p$ has much richer structure than ${\bf C}$ and offers
many new possibilities in related analysis and possible
applications.

It is often of interest for applications the ring of $p$-adic
integers ${\bf Z}_p = \{ x\in {\bf Q}_p: |x|_p \leq 1  \}$, {\it
i.e.} $p$-adic integers have the  representation $x = x_0 + x_1 p
+ x_2 p^2 + \cdots$.

$p$-Adic numbers can be suitably visualized by means of trees
\cite{holly} and as fractals in the Euclidean spaces
\cite{robert}. ${\bf Z}_p$  has property ${\bf Z}_p \supset {p\bf
Z}_p \supset {p^2\bf Z}_p \supset {p^3\bf Z}_p \cdots $ being  a
natural mathematical tool to investigate physical structures with
a hierarchy.

${\bf Z}_p$ is topologically compact and complete, and ${\bf Z}$
is dense in ${\bf Z}_p$. ${\bf Q}_p$ is locally compact, separable
and totally disconnected complete topological space. Some $p$-adic
spaces have rather exotic properties: $(i)$ all triangles are
isoceles and unequal side is the shortest one, $(ii)$ two discs
cannot partially intersect, and $(iii)$  every point of a disc may
be regarded as its center.

There is no natural ordering on ${\bf Q}_p$. However one can
introduce a linear order on ${\bf Q}_p$ in the following way: $x <
y$ if $| x |_p < | y|_p$, or if $| x|_p = | y|_p$ then there
exists such index $r \geq 0$ that digits satisfy $x_0 = y_0, x_{1}
= y_{1}, \cdots, x_{r-1} = y_{r-1}, x_{r} < y_{r}$. Here, $x_k$
and $y_k$ are digits related to $x$ and $y$, respectively, in
their expansions of the form $x =p^m (x_0 +x_1 p + x_2 p^2
\cdots)$. This ordering is very useful in calculation of $p$-adic
path integrals by the time discretization method.

There are primary two kinds of analyses on ${\bf Q}_p$ which are
of interest for physics, and they are based on two different
mappings: ${\bf Q}_p \to {\bf Q}_p$ and ${\bf Q}_p\to {\bf C}$. We
use both, in classical and quantum $p$-adic models, respectively.

Elementary $p$-adic valued functions and their derivatives are
defined by the same power series (i.e with the same rational
coefficients but $p$-adic arguments) as in the real case.   Their
regions of convergence are determined by means of $p$-adic norm
and they are usually restricted to $|x|_p < 1$. It is worth noting
that $\sum_{n\geq 0} P_k (n) n! x^n ,$ where $P_k (n)$ is a
polynomial in $n$ of degree $k$ with integer coefficients,
converges on $|x|_p \leq 1$ for all $p$  and their rational
summation is investigated in Ref. \cite{dragovich1}. As a definite
$p$-adic valued integral of an analytic function $f(x) = f_0 + f_1
x + f_2 x^2 + \cdots$ one takes difference of the corresponding
antiderivative in end points, {\it i.e.}
$$
 \int_a^b f(x) = \sum_{n=0}^\infty \frac{f_n}{n+1} \left( b^{n+1} -
 a^{n+1}   \right).                    \eqno(2.5)
$$

Usual complex-valued functions of $p$-adic variable, which are
employed in mathematical physics, are: ({\it i}) an additive
character $\chi_p (x) = \exp{2\pi i \{x \}_p}$, where $\{x \}_p$
is the fractional part of $x\in {\bf Q}_p$, ({\it ii}) a
multiplicative character $\pi_s (x) = |x|_p^s$, where $s\in {\bf
C}$, and ({\it iii}) locally constant functions with compact
support, like $\Omega (|x|_p)$, where
$$
\Omega (|x|_p) = \left\{  \begin{array}{ll}
                 1,   &   |x|_p \leq 1,  \\
                 0,   &   |x|_p > 1.
                 \end{array}    \right.                     \eqno(2.6)
$$
There is well defined Haar measure and integration
\cite{vladimirov1}. So, we have
$$
  \int_{{\bf Q}_p} \chi_p (ayx)\, dx = \delta_p (ay) = |a|^{-1}_p\,
  \delta_p (y) , \ \ a\neq 0,                                        \eqno(2.7)
$$
$$
  \int_{{\bf Q}_p} \chi_p (\alpha x^2 + \beta x) \, dx
  =\lambda_p (\alpha)\, |2\alpha|_p^{-\frac{1}{2}}
 \, \chi_p \left( -\frac{\beta^2}{4\alpha} \right), \ \ \alpha\neq
  0,
                                                   \eqno(2.8)
$$
where $\delta_p (u)$ is the $p$-adic Dirac $\delta$ function.  The
number-theoretic function $\lambda_p (x)$ in (2.8) is a map
$\lambda_p: {\bf Q}_p^\ast\to {\bf C}$ defined as follows:
$$
 \lambda_p (x) = \left\{   \begin{array}{ll}
                 1,    &  m=2j,  \ \ \ \ \ \ \  p\neq 2,  \\
                 \left(\frac{x_0}{p} \right),    &  m=2j+1,
 \ \  p\equiv 1(\mbox{mod}\ 4), \\
                 i\left(\frac{x_0}{p} \right),    &  m=2j+1,
  \ \  p\equiv 3(\mbox{mod}\ 4),
                 \end{array}   \right.            \eqno(2.9)
$$
$$
  \lambda_2 (x) = \left\{   \begin{array}{ll}
                  \frac{1}{\sqrt 2} [1 + (-1)^{x_{1}} i],
   &  m= 2j,  \\
                 \frac{1}{\sqrt 2} (-1)^{x_{1}+ x_{2}}
                 [1 + (-1)^{x_{1}} i],      &   m=2j+1,
                 \end{array}
                 \right.        \eqno(2.10)
$$
where $x$ is presented as $x= p^m \ (x_0 + x_1 p + x_2 p^2
+\cdots)$, and $m, j\in {\bf Z}.\ $
 $\left(\frac{x_0}{p} \right)$ is the  Legendre symbol defined as
$$
 \left(\frac{a}{p} \right) = \left\{   \begin{array}{ll}
                 1,    &  \mbox{if} \ \ \ a\equiv y^2(\mbox{mod}\ p) ,  \\
                -1,    &  \mbox{if} \ \ \ a\not\equiv y^2(\mbox{mod}\ p) ,\\
                 0,    &   \mbox{if} \ \ \  a\equiv 0(\mbox{mod}\ p),
                 \end{array}   \right.            \eqno(2.11)
$$
and ${\bf Q}_p^\ast = {\bf Q}_p\setminus \{0\}$. It is often
sufficient to use  their standard  properties:
$$
  \lambda_p (a^2 x)= \lambda_p (x), \ \  \lambda_p (x) \lambda_p (-x) =1,
\ \  \lambda_p (x)\lambda_p (y)  = \lambda_p (x+y)  \lambda_p
(x^{-1}+y^{-1}),
$$
$$
|\lambda_p (x)|_\infty = 1,   \ \ a\neq 0.      \eqno(2.12)
$$

Recall that the real analogues of  (2.7) and (2.8) have the same
form, {\it i.e.}
$$
  \int_{{\bf Q}_\infty} \chi_\infty (ayx)\, dx = \delta_\infty (ay)
  = |a|^{-1}_\infty\,
  \delta_\infty (y) , \ \ a\neq 0,                                        \eqno(2.13)
$$
$$
  \int_{Q_\infty} \chi_\infty (\alpha x^2 + \beta x)\, dx
  =\lambda_\infty (\alpha)\, |2\alpha|_\infty^{-\frac{1}{2}}
 \, \chi_\infty \left( -\frac{\beta^2}{4\alpha} \right), \ \ \alpha\neq 0,
                                                   \eqno(2.14)
$$
where ${\bf Q_\infty}\equiv {\bf R}$, $\, \, \chi_\infty (x) =
\exp{(-2\pi i x)}$ is additive character in the real case and
$\delta_\infty$ is the ordinary Dirac $\delta$ function. Function
$\lambda_\infty (x)$ is defined as
$$
  \lambda_\infty (x) = \sqrt{\frac{\mbox{sign}\ x}{i}}, \ \
  x\in {\bf R}^\ast ={\bf R}\setminus \{ 0 \} \eqno(2.15)
$$
and exhibits the same properties (2.12).

For a more  information on  usual properties of $p$-adic numbers
and related analysis one can see
\cite{gouvea,schikhof,vladimirov1,mahler,koblitz,gelfand}.

\section{Adeles and Their Functions}

Real and $p$-adic numbers are unified in the form of adeles. An
adele $x$ \cite{gelfand,weil,platonov} is an infinite sequence
$$
  x= (x_\infty, x_2, \cdots, x_p, \cdots),             \eqno(3.1)
$$
where $x_\infty \in {\bf R}$ and $x_p \in {\bf Q}_p$ with the
restriction that for all but a finite set $\bf S$ of primes $p$
one has  $x_p \in {\bf Z}_p$.   Rational numbers are naturally
embedded in the space of adeles.   Componentwise addition and
multiplication are ordinary arithmetical  operations on the ring
of adeles ${\cal A}$, which can be regarded as
$$
 {\cal A} = \bigcup_{{\bf S}} {\cal A} ({\bf S}),
 \ \  {\cal A}({\bf S}) = {\bf R}\times \prod_{p\in {\bf S}} {\bf Q}_p
 \times \prod_{p\not\in {\bf S}} {\bf Z}_p.         \eqno(3.2)
$$
${\cal A}$ is a locally compact topological space.

There are also two kinds of analyses over topological ring of
adeles ${\cal A}$, which are generalizations of the corresponding
analyses over $\bf R$ and  ${\bf Q}_p$. The first one is related
to the mapping ${\cal A}\to {\cal A}$ and the other one to ${\cal
A}\to {\bf C}$. In complex-valued adelic analysis it is worth
mentioning  an additive character
$$
 \chi (x) = \chi_\infty (x_\infty) \prod_p \chi_p (x_p),   \eqno(3.3)
$$
a multiplicative character
$$
  |x|^s = |x_\infty|_\infty^s \prod_p |x_p|_p^s, \ \ s\in {\bf C},
                                                    \eqno(3.4)
$$
and elementary functions of the form
$$
 \phi (x) = \phi_\infty (x_\infty) \prod_{p\in {\bf S}} \phi_p (x_p)
 \prod_{p\not\in {\bf S}} \Omega (|x_p|_p),           \eqno(3.5)
$$
where $\phi_\infty (x_\infty)$ is an infinitely differentiable
function on ${\bf R}$ such that $|x_\infty |_\infty^n \phi_\infty
(x_\infty) \to 0$ as $|x_\infty|_\infty \to \infty$ for any $n\in
\{0,1,2,\cdots  \}$, and $\phi_p (x_p)$ are some locally constant
functions with compact support. All finite linear combinations of
elementary functions (3.5) make the set $S({\cal A})$ of the
Schwartz-Bruhat adelic functions. The Fourier transform of $\phi
(x)\in S({\cal A})$, which maps $S(\cal A)$ onto ${\cal A}$, is
$$
 \tilde{\phi}(y) = \int_{\cal A} \phi (x)\chi (xy)dx,      \eqno(3.6)
$$
where $\chi (xy)$ is defined by (3.3) and $dx = dx_\infty dx_2
dx_3 \cdots$ is the Haar measure on ${\cal A}$.

One can define the Hilbert space on ${\cal A}$, which we will
denote by $L_2({\cal A})$. It contains infinitely many
complex-valued functions of adelic argument (for example,
$\Psi_1(x), \Psi_2(x), \cdots$) with scalar product
$$
 (\Psi_1,\Psi_2) = \int_{\cal A} \bar{\Psi}_1(x) \Psi_2(x) dx
$$
and norm
$$
  ||\Psi|| = (\Psi,\Psi)^{\frac{1}{2}} < \infty .
$$
 A basis of $L_2({\cal A})$ may be given by the set of
orthonormal eigefunctions in a spectral problem of the evolution
operator $U (t)$, where $t\in {\cal A}$. Such eigenfunctions have
the form
$$
\psi_{{\bf S},\alpha} (x,t) = \psi_n^{(\infty)}(x_\infty,t_\infty)
 \prod_{p\in {\bf S}} \psi_{\alpha_p}^{(p)} (x_p,t_p)
 \prod_{p\not\in {\bf S}} \Omega (|x_p|_p),                    \eqno(3.7)
$$
where $\psi_n^{(\infty)} \in L_2({\bf R})$ and
$\psi_{\alpha_p}^{(p)} \in L_2({\bf Q}_p) $ are eigenfunctions in
ordinary and $p$-adic cases, respectively. Indices $n, \alpha_2,
\cdots, \alpha_p, \cdots$ are related to the corresponding real
and $p$-adic eigenvalues of the same observable in a physical
system. $\Omega (|x_p|_p)$ is an element of $L_2({\bf Q}_p)$,
defined by (2.6), which is invariant under transformation of an
evolution operator $U_p(t_p)$ and provides convergence of the
infinite product (3.7). For a fixed ${\bf S}$, states  $\psi_{{\bf
S},\alpha} (x,t)$ in (3.7) are eigefunctions of   $L_2({\cal
A}({\bf S}))$, where ${\cal A}({\bf S})$ is a subset of adeles
${\cal A}$ defined by (3.2). Elements of $L_2({\cal A})$ may be
regarded as superpositions $\Psi (x) = \sum_{{\bf S}} C({\bf S})
\Psi_{{\bf S}}(x), $ where $\Psi_{{\bf S}}(x)\in L_2({\cal A}({\bf
S}))$, {\it  i.e.}
$$
\Psi_{{\bf S}}(x) = \Psi_\infty(x_\infty)\prod_{p\in {\bf
S}}\Psi_p(x_p) \prod_{p\not\in {\bf S}}\Omega(| x_p |) , \quad
x\in {\cal A},                \eqno(3.8)
$$
and  $\sum_{{\bf S}} |C({\bf S})|_\infty^2 =1$.

Theory of $p$-adic generalized functions is presented in Ref.
\cite{vladimirov1}. Construction of generalized functions on
adelic spaces is a hard task, but there is some progress within
adelic quantum mechanics \cite{dragovich1a}.

\section{Quantum Mechanics on Adelic Spaces}

There are a number of reasons to use $p$-adic numbers and adeles
in investigation of mathematical and theoretical aspects of modern
quantum physics. Some primary of them are: ({\it i}) the field of
rational numbers ${\bf Q}$, which contains all observational and
experimental numerical data, is a dense subfield not only in ${\bf
R}$ but also in the fields of $p$-adic numbers ${\bf Q}_p$, ({\it
ii}) there is an  analysis \cite{vladimirov1} within and over
${\bf Q}_p$ like that one related to ${\bf R}$, ({\it iii})
general mathematical methods and fundamental physical laws should
be invariant \cite{volovich1} under an interchange of the number
fields ${\bf R}$ and ${\bf Q}_p$, ({\it iv}) there is a quantum
gravity uncertainty  $\Delta x$ (see (1.1)), when measures
distances around the Planck length $\ell_0$, which restricts
priority of Archimedean geometry based on real numbers and gives
rise to employment of non-Archimedean geometry related to $p$-adic
numbers, ({\it v}) it seems to be quite reasonable to extend
standard Feynman's  path integral method to non-Archimedean
spaces, and ({\it vi}) adelic quantum mechanics \cite{dragovich2}
is consistent with all the above assertions.

In order to investigate adelic quantum theory in a systematic way
it is natural to start by formulation of adelic quantum mechanics.
According to \cite{dragovich2}, adelic quantum mechanics contains
three main ingredients $ (L_2 ({\cal A}), W(z), U(t)) $ where:
$(i)$ $L_2 ({\cal A})$ is an adelic Hilbert space, $(ii)$ $W(z)$
denotes Weyl quantization of complex-valued functions on adelic
classical phase space, and $(iii)$  $U(t)$  is  the unitary
representation of an evolution operator on $L_2 ({\cal A})$.

Canonical noncommutativity in $p$-adic case can be represented by
the Weyl operators ($h=1$)
$$
\hat Q_p(\alpha) \psi_p(x)=\chi_p(\alpha x)\psi_p(x) \eqno(4.1)
$$
$$
\hat K_p(\beta)\psi(x)=\psi_p(x+\beta)  \eqno(4.2)
$$
in the form
$$
\hat Q_p(\alpha)\hat K_p(\beta)=\chi_p(\alpha\beta) \hat
K_p(\beta)\hat Q_p(\alpha). \eqno(4.3)
$$
 It is possible to introduce the product of unitary operators
$$
\hat W_p(z)=\chi_p(-\frac 1 2 qk)\hat K_p(\beta)\hat Q_p(\alpha),
\quad z\in {\bf Q}_p\times {\bf Q}_p, \eqno(4.4)
$$
which is a unitary representation of the Heisenberg-Weyl group.
Recall that this group consists of the elements $(z,\alpha)$ with
the group product
$$
(z,\alpha)\cdot (z',\alpha ')=(z+z',\alpha+\alpha '+\frac{1}{2}
B(z,z')), \eqno(4.5)
$$
where $B(z,z') = -kq'+qk'$ is a skew-symmetric bilinear form on
the phase space.

Dynamics of a $p$-adic quantum model is described by a unitary
evolution operator $U_p(t)$ in terms of its kernel ${\cal
K}_t^{(p)}(x,y)$
$$
 U_p(t)\psi^{(p)}(x)=\int_{{\bf Q}_p}{\cal K}^{(p)}_t(x,y)\psi^{(p)}(y) dy. \eqno(4.6)
$$
In this way, $p$-adic quantum mechanics \cite{vladimirov2} is
given by a triple
$$
(L_2(Q_p), W_p(z_p), U_p(t_p)). \eqno(4.7)
$$
Keeping in mind that ordinary quantum mechanics can be also given
as the analogue of (4.7), ordinary and $p$-adic quantum mechanics
can be unified in the form of the above-mentioned adelic quantum
mechanics \cite{dragovich2}.

Adelic evolution operator $U(t)$ is defined by
$$
U(t)\psi(x)=\int_{{\cal A}} {\cal
K}_t(x,y)\psi(y)dy=\prod\limits_{v}{} \int_{Q_{v}}{\cal
K}_{t}^{(v)}(x_{v},y_{v})\psi^{(v)}(y_v) dy_{v}. \eqno(4.8)
$$
where $v=\infty, 2, 3,\cdots, p,\cdots$. The eigenvalue problem
for $U(t)$ reads
$$
U(t)\psi _{\alpha } (x)=\chi (E_{\alpha} t) \psi _{\alpha} (x),
\eqno(4.9)
$$
where $\psi_{\alpha }$ are adelic eigenfunctions, $E_{\alpha
}=(E_{\infty}, E_{2},..., E_{p},...)$ is the corresponding adelic
energy. Note that any adelic eigenfunction has the form (3.7).

A suitable way to compute $p$-adic propagator ${\cal K}_p
(x'',t'';x',t')$ is to use Feynman's path integral method, {\it
i.e.}
$$
{\cal K}_p(x'',t'';x',t') = \int_{x',t'}^{x'',t''} \chi_p \left(
-\frac{1}{h} \int_{t'}^{t''} L(\dot{q},q,t) dt  \right) {\cal D}q.
\eqno(4.10)
$$
It has been evaluated \cite{dragovich3,dragovich4}  for quadratic
Lagrangians in the same way for real and $p$-adic cases, and the
following exact general expression for propagator is obtained:
$$
 {\cal K}_v(x'',t'';x',t')= \lambda_v \left( - \frac{1}{2h}
\frac{\partial^2{\bar S}}{\partial x''\partial x'} \right) \left|
\frac{1}{h}\frac{\partial^2{\bar S}}{\partial x''\partial x'}
\right|_v^{\frac{1}{2}} \chi_v(-\frac{1}{h} {\bar S}
(x'',t'';x',t')),                    \eqno(4.11)
$$
where $\lambda_v$ functions satisfy relations (2.12) and ${\bar S}
(x'',t'';x',t'))$ is the action for classical trajectory. When one
has a system with more than one dimension with uncoupled spatial
coordinates, then the total $v$-adic propagator is the product of
the corresponding one-dimensional propagators.

As an illustration of $p$-adic and adelic quantum-mechanical
models, the following one-dimensional systems with the quadratic
Lagrangians were considered: a free particle and a harmonic
oscillator \cite{vladimirov1,dragovich2}, a particle in a constant
field \cite{dragovich5}, a free relativistic particle
\cite{dragovich6} and a harmonic oscillator with time-dependent
frequency \cite{dragovich7}.

Adelic quantum mechanics takes into account ordinary as well as
$p$-adic quantum effects and may be regarded as a starting point
for construction of a more complete quantum cosmology and
string/M-theory. In the low-energy limit adelic quantum mechanics
effectively becomes the ordinary one \cite{dragovich6}.

\section{Adelic Quantum Cosmology}

The main task of quantum cosmology is to describe  the very early
stage in the evolution of the Universe. At this stage, the
Universe was in a quantum state, which should be described by a
wave function. Usually one takes it that this wave function is
complex-valued and depends on some real parameters. Since quantum
cosmology is related to the Planck scale phenomena it is natural
to reconsider its foundations. We  maintain here the standard
point of view that the wave function takes complex values, but we
treat its arguments in a more complete way. Namely, we  regard
space-time coordinates, gravitational and matter fields to be
adelic, {\it i.e.} they have real as well as $p$-adic properties
simultaneously.

As there is no appropriate complex-valued $p$-adic Schr\"odinger
equation, so there is not also $p$-adic generalization of the
Wheeler - De Witt equation for cosmological models. Instead of
differential approach, Feynman's path integral method was
exploited \cite{dragovich8} and minisuperspace cosmological models
are investigated as models of adelic quantum mechanics
\cite{dragovich9,dragovich10}.

Adelic minisuperspace quantum cosmology is an application of
adelic quantum mechanics to the  cosmological models. In the path
integral approach to standard quantum cosmology, the starting
point is Feynman's path integral method, {\it i.e.} the amplitude
to go from one state with intrinsic metric $h_{ij}'$ and matter
configuration $\phi'$ on an initial hypersurface $\Sigma'$ to
another state with metric $h_{ij}''$ and matter configuration
$\phi''$ on a final hypersurface $\Sigma''$ is given by the path
integral
$$
{\cal K}_\infty ( h_{ij}'',\phi'',\Sigma''; h_{ij}',\phi',\Sigma')
= \int {\cal D}{(g_{\mu\nu})}_\infty {\cal D}(\Phi)_\infty \
\chi_\infty(-S_\infty[g_{\mu\nu},\Phi]), \eqno(5.1)
$$
over all four-geometries $g_{\mu\nu}$ and matter configurations
$\Phi$, which interpolate between the initial and final
configurations. In (5.1) $S_\infty [g_{\mu\nu},\Phi]$ is an
Einstein-Hilbert action for the gravitational and matter fields.
This action can be calculated using metric in the standard 3+1
decomposition
$$
ds^2=g_{\mu\nu}dx^\mu dx^\nu=-(N^2 -N_i N^i)dt^2 + 2N_i dx^i dt +
h_{ij} dx^i dx^j, \eqno(5.2)
$$
where $N$ and $N_i$ are the lapse and shift functions,
respectively. To perform $p$-adic and adelic generalization we
make first $p$-adic counterpart of the action using
form-invariance under change of real to the $p$-adic number
fields. Then we generalize (5.1) and introduce $p$-adic
complex-valued cosmological amplitude
$$
 {\cal K}_p ( h_{ij}'',\phi'',\Sigma''; h_{ij}',\phi',\Sigma') =
 \int{\cal D}{(g_{\mu\nu})}_p{\cal D}(\Phi)_p \
\chi_p(-S_p[g_{\mu\nu},\Phi]). \eqno(5.3)
$$

Since the space of all three-metrics and matter field
configurations on a three-surface, called superspace,  has
infinitely many dimensions, in computation one takes an
approximation. A useful approximation is to truncate the infinite
degrees of freedom to a finite number $q_\alpha(t)$,
($\alpha=1,2,...,n$). In this way, one obtains a  minisuperspace
model. Usually, one restricts the four-metric to be of the form
(5.2), with $N^i=0$ and $h_{ij}$ approximated by $q_\alpha(t)$.
For the homogeneous and isotropic cosmologies  the  metric is a
Robertson-Walker one, which spatial sector reads
$$
h_{ij}dx^idx^j=a^2(t)d\Omega_3^2 = a^2(t) \left[
d\chi^2+\sin^2\chi(d\theta^2+\sin^2\theta d\varphi^2) \right],
\eqno(5.4)
$$
where $a(t)$ is a scale factor. If we use also a single scalar
field $\phi$, as a matter content of the model, minisuperspace
coordinates are $ a $ and $\phi$. More generally, models can be
homogeneous but also anisotropic ones.

For the boundary condition $q_\alpha(t'')=q_\alpha''$, \
$q_\alpha(t')=q_\alpha'$ in the gauge $ N=1$, we have $v$-adic
minisuperspace propagator
$$
 {\cal K}_v (q_{\alpha}''|q_{\alpha}') =\int dt \ {\cal
K}_v (q_{\alpha}'',t'';q_{\alpha}',t'), \quad t=t''-t' ,
\eqno(5.5)
$$
where
$$
 {\cal K}_v (q_{\alpha}'',t'';q_{\alpha}',t') =\int {\cal
D}q_\alpha \ \chi_v(-S_v[q_\alpha]),   \eqno(5.6)
$$
is an ordinary quantum-mechanical propagator between fixed
minisuperspace coordinates ($q_\alpha',q_\alpha''$)  in  fixed
times. $S_v$ is the $v$-adic action of the minisuperspace model,
{\it i.e.}
$$
S_v[q_\alpha]= \int_{t'}^{t''} dt  \left[ \frac{1}{2}
f_{\alpha\beta}(q)\dot q^\alpha\dot q^\beta-U(q) \right],
\eqno(5.7)
$$
where $f_{\alpha\beta}$ is a minisuperspace metric
$(ds^2_m=f_{\alpha\beta}dq^\alpha dq^\beta)$ with an indefinite
signature ($-,+,+,\dots$). This metric includes spatial
(gravitational) components and also matter variables for the given
model.

The standard minisuperspace ground-state wave function in the
Hartle-Hawking (no-boundary) proposal \cite{hartle} is defined by
functional integration in the Euclidean version of
$$
\Psi_\infty[h_{ij}]= \int{\cal D}(g_{\mu\nu})_\infty {\cal
D}(\Phi)_\infty \ \chi_\infty(-S_\infty[g_{\mu\nu},\Phi]),
\eqno(5.8)
$$
over all compact four-geometries $g_{\mu\nu}$ which induce
$h_{ij}$ at the compact three-manifold. This three-manifold is the
only boundary of the all four-manifolds. Extending Hartle-Hawking
proposal to the $p$-adic minisuperspace,  an adelic Hartle-Hawking
wave function is the infinite product
$$
\Psi[h_{ij}]= \prod_{v}\int{\cal D}(g_{\mu\nu})_v {\cal D}(\Phi)_v
\ \chi_\upsilon(-S_v [g_{\mu\nu},\Phi]), \eqno(5.9)
$$
where path integration must be performed over both, Archimedean
and non-Archimedean geometries. If  evaluation of the
corresponding functional integrals for a minisuperspace model
yields $\Psi(q_\alpha)$ in the form (3.7), then we  say that such
cosmological model is a Hartle-Hawking adelic one. It is shown
\cite{dragovich8} that the de Sitter minisuperspace model in $D=4$
space-time dimensions is  the Hartle-Hawking adelic one.

It is shown in \cite{dragovich9,dragovich10}   that $p$-adic and
adelic generalization of the minisuperspace cosmological models
can be successfully performed in the framework of $p$-adic and
adelic quantum mechanics \cite{dragovich2} without use of the
Hartle-Hawking approach.  The following cosmological models are
investigated: the de Sitter model, model with a homogeneous scalar
field, anisotropic Bianchi model with three scale factors and some
two-dimensional minisuperspace models. As a result of $p$-adic
effects and adelic approach, in these models there is some
discreteness of minisuperspace and cosmological constant. This
kind of discreteness was obtained for the first time in the
context of  the Hartle-Hawking adelic de Sitter quantum model
\cite{dragovich8}.

\section{Adelic String/M-theory}

A notion of $p$-adic string was introduced in \cite{volovich2},
where the hypothesis on the existence of non-Archimedean geometry
at the Planck scale was made, and string theory with $p$-adic
numbers was initiated. In particular, generalization of the usual
Veneziano and Virasoro-Shapiro amplitudes with complex-valued
multiplicative characters over various number fields was proposed
and  $p$-adic valued Veneziano amplitude was constructed by means
of $p$-adic interpolation. Very successful  $p$-adic analogues of
the Veneziano and Virasoro-Shapiro amplitudes were proposed in
\cite{freund1} as the corresponding Gel'fand-Graev \cite{gelfand}
beta functions. Using this approach, Freund and Witten obtained
\cite{freund2} an attractive adelic formula, which states that the
product of the crossing symmetric Veneziano (or Virasoro-Shapiro)
amplitude and its all $p$-adic counterparts equals unit (or a
definite constant). This gives possibility to consider an ordinary
four-point function, which is rather complicate, as an infinite
product of its inverse $p$-adic analogues, which have simple
forms. These first papers induced an interest in various aspects
of $p$-adic string theory (for a review, see
\cite{freund3,vladimirov1}). A recent interest in $p$-adic string
theory has been mainly related to the generalized adelic formulas
for four-point string amplitudes \cite{vladimirov3}, the tachyon
condensation \cite{sen}, nonlinear dynamics \cite{vladimirov4} and
the new attractive  adelic approach \cite{dragovich11}.

Like in the ordinary string theory, the starting point in $p$-adic
string theory is a construction of the corresponding scattering
amplitudes.  Recall that the ordinary crossing symmetric Veneziano
amplitude can be presented in the following forms:
$A_\infty(k_1,\cdots,k_4) \equiv   A_\infty(a,b)$
$$
= g^2 \int_{{\bf R}} \vert x \vert_\infty^{a-1}
  \vert 1-x\vert_\infty^{b-1} dx
     \eqno(6.1)
$$
$$
 = g^2 \left[\frac{\Gamma{(a)} \Gamma{(b)} }{\Gamma{(a+b)}} +
 \frac{\Gamma{(b)}\Gamma{(c)}}{\Gamma{(b+c)}} +
 \frac{\Gamma{(c)}\Gamma{(a)}}{\Gamma{(c+a)}}\right]     \eqno(6.2)
$$
$$
 = g^2 \frac{\zeta(1-a)}{\zeta(a)} \frac{\zeta(1-b)}{\zeta(b)}
 \frac{\zeta(1-c)}{\zeta(c)}                     \eqno(6.3)
$$
$$
=g^2 \int {\cal D}X \exp\left(  -\frac{i}{2\pi}
 \int d^2 \sigma \partial^\alpha X_\mu \partial_\alpha X^\mu \right)
 \prod_{j=1}^4 \int d^2 \sigma_j \exp\left( i k_\mu^{(j)} X^\mu
\right) ,                           \eqno(6.4)
$$
where $\hbar=1,\ T=1/\pi$, and $a=-\alpha (s) = - 1 -\frac{s}{2},
\ b=-\alpha (t), \ c=-\alpha (u)$ with the condition $s+t+u = -8$,
i.e. $a+b+c=1$.

To introduce a $p$-adic Veneziano amplitude one can consider
$p$-adic analogues of all the above four expressions. $p$-Adic
generalization of the first expression was proposed in
\cite{freund1} and it reads
$$
 A_p (a,b) = g_p^2 \int_{{\bf Q}_p} \vert x\vert_p^{a-1}
 \vert 1-x\vert_p^{b-1} dx ,                 \eqno(6.5)
$$
where $\vert \cdot \vert_p$ denotes $p$-adic absolute value.
 In this case only
string world-sheet parameter $x$ is treated as $p$-adic variable,
and all other quantities maintain their usual (real) valuation. An
attractive adelic formula of the form
$$
 A_\infty (a,b) \prod_p A_p (a,b) =1            \eqno(6.6)
$$
was found \cite{freund2}, where $A_\infty (a,b)$ denotes the usual
Veneziano amplitude (6.1). A similar product formula holds also
for the Virasoro-Shapiro amplitude. These infinite products are
divergent, but they can be successfully regularized.
Unfortunately, there is a problem to extend this formula to the
higher-point functions.

$p$-Adic analogues of (6.2) and (6.3) were also proposed in
\cite{volovich2} and \cite{dragovich12}, respectively. In these
cases, world-sheet, string momenta and amplitudes are manifestly
$p$-adic. Since string amplitudes are $p$-adic valued functions,
it is not so far enough clear their physical interpretation.

Expression (6.4) is based on Feynman's path integral method, which
is generic for all quantum systems and has successful $p$-adic
generalization.  $p$-Adic counterpart of (6.4) is proposed in
\cite{dragovich11} and has been partially elaborated in
\cite{dragovich13} and \cite{dragovich14}.  Note that in this
approach, $p$-adic string amplitude is complex-valued, while not
only the world-sheet parameters but also target space coordinates
and string momenta are $p$-adic variables. Such $p$-adic
generalization is a natural extension of the formalism of $p$-adic
\cite{vladimirov2} and adelic \cite{dragovich2} quantum mechanics
to string theory. This is a promising subject and should be
investigated in detail, and applied to the branes and M-theory,
which is presently the best candidate for the fundamental physical
theory.

\section{Concluding Remarks}

One of the very interesting and fruitful recent developments in
string theory  has been noncommutative geometry and the
corresponding noncommutative field theory, which  may be regarded
as a deformation of the ordinary one in which field multiplication
is replaced by the Moyal (star) product
$$
 (f\star g)(x) =\exp\left[\frac{i \hbar}{2} \theta^{ij}\frac{
\partial}{\partial y^{i}} \frac{
\partial}{\partial z^j}\right] f(y)g(z)\vert_{y=z =x},     \eqno(7.1)
$$
where $x= (x^1,x^2,\cdots,x^d)$ is a spatial point  , and
$\theta^{ij} =-\theta^{ji}$ are noncommutativity parameters.
Replacing the ordinary product between noncommutative coordinates
by the Moyal product (7.1) we have
$$
 x^{i}\star x^j - x^j \star x^{i} = i\hbar \theta^{ij} ,  \eqno(7.2)
$$
which resembles the usual Heisenberg algebra. It is worth noting
that one can introduce \cite{dragovich14} the Moyal product in
$p$-adic quantum mechanics and it reads
$$
( f \ast  g)(x)=\int_{{\bf Q}_p^d}\int_{{\bf Q}_p^d} dk dk' \
\chi_p(-(x^ik_i+x^jk'_j)+\frac{1}{2} k_ik'_j\theta^{ij})\tilde
f(k)\tilde g(k'), \eqno(7.3)
$$
where $d$ denotes  spatial dimensionality, and $\tilde f(k)$, \
$\tilde g(k')$ denote the Fourier transforms of $f(x)$ and $g(x)$.
Some real, $p$-adic and adelic aspects of the noncommutative
scalar solitons \cite{strominger} are investigated in Ref.
\cite{dragovich15}.

A natural extension of adelic quantum mechanics to adelic field
theory is  considered \cite{dragovich16}, as well.  $p$-Adic
quantum mechanics with $p$-adic valued wave functions has been
investigated and  presented in
\cite{vladimirov1,khrennikov1,khrennikov2}. There have been also
applications of $p$-adic numbers and adeles relevant for some
other directions of mathematical research in physics, but I
restricted this review to some aspects of quantum theory.

\bigskip

{\bf Acknowledgements.\ } The work on this paper was supported in
part by the Serbian Ministry of Science, Technologies and
Development under contract No 1426 and by RFFI grant 02--01-01084
.


\begin{thebibliography}{99} 


\bibitem{schikhof} W.H. Schikhof, {\it Ultrametric Calculus : an introduction to $p$-adic analysis},
             Cambridge U.P., Cambridge, 1984.
\bibitem{holly} J.E. Holly, Pictures of Ultrametric Spaces, the $p$-adic Numbers, and Valued
               Fields, {\it Amer. Math. Monthly} {\bf 108} (2001) 721-728.
\bibitem{robert} A. Robert, Euclidean models of $p$-adic spaces, {\it Lecture Notes in Pure and Applied Mathematics}
              {\bf 192} (1997) 349-361.
\bibitem{dragovich1} B. Dragovich, On some $p$-adic series with
                     factorials, {\it Lecture Notes in Pure and Applied Mathematics}
                     {\bf 192} (1997) 95-105.
\bibitem{vladimirov1} V.S. Vladimirov, I.V. Volovich and E.I. Zelenov,
                {\it $p$-Adic Analysis and Mathematical Physics}, World
                 Scientific, Singapore, 1994.
\bibitem{gouvea} F.Q. Gouvea, {\it $p$-adic Numbers : An
                              introduction}, Universitext,
                              Springer-Verlag, 1993.
\bibitem{mahler} K. Mahler, {\it $p$-adic numbers and their
                 functions}, Cambridge tracts in mathematics {\bf 76}, Cambridge
                 U.P., Cambridge, 1980.
\bibitem{koblitz} N. Koblitz, {\it $p$-adic numbers, $p$-adic analysis and zeta
                  functions}, London Mathematical Society Lecture Notes Series {\bf
                  46}, Cambridge U.P., Cambridge, 1980.
\bibitem{gelfand} I.M. Gel'fand, M.I. Graev and I.I. Piatetskii-Shapiro,
                 {\it Representation Theory and Automorphic Functions} (in Russian), Nauka,
                 Moscow, 1966.
\bibitem{weil}  A. Weil, {\it Adeles and Algebraic Geometry},
                 Progress in Mathematics  {\bf 23}, Birkh\"auser, 1982.
\bibitem{platonov} V.P. Platonov and A.S. Rapinchuk, {\it Algebraic
               Groups and Number Theory} (in Russian), Nauka, Moscow, 1991.
\bibitem{dragovich1a} B. Dragovich, On Generalized Functions in
                 Adelic Quantum Mechanics, {\it Integral Transforms and Special
                   Functions} {\bf 6} (1998) 197-203.
\bibitem{volovich1} I.V. Volovich, {\it Number theory as the ultimate physical theory},
        CERN preprint, CERN-TH.4781/87 (July 1987).
\bibitem{dragovich2} B. Dragovich, Adelic Model of Harmonic
                     Oscillator,
                     {\it Theor. Math. Phys.} {\bf 101} (1994) 1404-1412;
                 Adelic Harmonic Oscillator, {\it Int. J. Mod. Phys.} {\bf A10} (1995) 2349-2365.
\bibitem{vladimirov2} V.S. Vladimirov and I.V. Volovich, $p$-Adic Quantum
                      Mechanics,
                 {\it Commun. Math. Phys.} {\bf 123} (1989) 659-676.
\bibitem{dragovich3} G.S. Djordjevi\'c and B. Dragovich, $p$-Adic Path Integrals for
                 Quadratic Actions, {\it Mod. Phys.
                 Lett.} {\bf A 12} (1997) 1455-1463.
\bibitem{dragovich4} G.S. Djordjevi\'c, B. Dragovich and Lj. Ne\v si\'c,
        Adelic Path Integrals  for Quadratic Lagrangians, to be
        published in {\it Infinite Dimensinal Analysis, Quantum Probability and Related
        Topics},
        hep-th/0105030.
\bibitem{dragovich5} G.S. Djordjevic and B. Dragovich, On $p$-adic functional
        integration,   {\it Proc. of the II Mathematical Conference},
        Pri\v stina, Yugoslavia (1997) 221-228.
\bibitem{dragovich6} G.S. Djordjevi\'c, B. Dragovich, Lj. Ne\v si\'c,
        $p$-Adic and Adelic Free Relativistic Particle,
        {\it Mod. Phys. Lett.} {\bf A 14} (1999) 317-325.
\bibitem{dragovich7} G.S. Djordjevic and B. Dragovich, $p$-Adic and Adelic Harmonic Oscillator
        with a Time-dependent Frequency, {\it Theor. Math. Phys.}
        {\bf 124} (2000) 1059-1067.
\bibitem{dragovich8} B. Dragovich,  Adelic Wave Function of the Universe,
        {\it Proc. of the Third A. Friedmann Int. Seminar on Gravitation and
        Cosmology}, Friedmann Lab. Publishing, St. Petersburg, 1995,
        pp. 311-321.
\bibitem{dragovich9} B. Dragovich and Lj. Nesic, $p$-Adic and Adelic Generalization of
         Quantum Cosmology, {\it Gravitation and Cosmology} {\bf 5} (1999)
                 222-228.
\bibitem{dragovich10} G.S. Djordjevic, B. Dragovich, Lj. Nesic and I.V.
            Volovich, $p$-Adic and Adelic Minisuperspace Quantum Cosmology,
            {\it Int. J. Mod. Phys.} {\bf  A17} (2002) 1413-1433.
\bibitem{hartle} J.B. Hartle and S.W. Hawking, Wave function of the Universe, {\it Phys. Rev. }
        {\bf 28} (1983) 2960-2075.
\bibitem{volovich2} I.V. Volovich, $p$-Adic string, {\it Class. Quantum Grav.} {\bf 4} (1987) L83-L87.
\bibitem{freund1} P.G.O. Freund and M. Olson, Non-Archimedean strings, {\it Phys. Lett.}
           {\bf B 199} (1987) 186-190.
\bibitem{freund2} P.G.O. Freund and E. Witten, Adelic string amplitudes, {\it Phys. Lett.}
              {\bf B 199} (1987) 191-194.
\bibitem{freund3} L. Brekke and P.G.O. Freund, $p$-Adic numbers in Physics, {\it Phys. Rep.} {\bf
233} (1993) 1-63.
\bibitem{vladimirov3} V.S. Vladimirov,  Adelic Formulas for Gamma and
        Beta Functions of One-Class Quadratic Fields: Applications to
        4-Particle Scattering String Amplitudes,  {\it Proc. Steklov Math.
         Institute} {\bf 228} (2000) 67-73 ,   math-ph/0004017.
\bibitem{sen} D. Ghoshal and A. Sen,  Tachyon Condesation and
       Brane Descent Relations in $p$-adic String Theory, {\it Nucl.
       Phys.} B{\bf 584} (2000) 300,  hep-th/0003278.
\bibitem{vladimirov4} V.S. Vladimirov and Ya.I. Volovich, On the
       Nonlinear Dynamical Equation in the $p$-adic String Theory,
       math-ph/0306018.
\bibitem{dragovich11} B. Dragovich,  On Adelic Strings, hep-th/0005200.
\bibitem{dragovich12} I.Ya. Aref'eva, B. Dragovich and I.V. Volovich, On the adelic string amplitudes,
         {\it Phys. Lett.} {\bf B 209 } (1988) 445-450.
\bibitem{dragovich13} B. Dragovich, $p$-Adic and Adelic Strings,
         {\it Proc. Int. Conference dedicated to the memory of Prof. E.
         Fradkin: Quantization, Gauge Theory and Strings}, Scientific
         World, Moscow, 2001, pp. 108-114.
\bibitem{dragovich14} B. Dragovich and I.V. Volovich, $p$-Adic
         Strings and Noncommutativity, {\it Proc.  Workshop on
         Noncommutative Structures in Mathematics and Physics}, NATO
         Science Series: II. Mathematics, Physics and Chemistry - Vol. 22.,
         Kluwer AP, 2001, pp. 391-399.
\bibitem{strominger} R. Gopakumar, S. Minwalla and A. Strominger,
          Noncommutative Solitons, {\it JHEP} {\bf
           05} : 020, hep-th/003160.
\bibitem{dragovich15} B. Dragovich and B. Sazdovic, Real, $p$-Adic
and Adelic Scalar Solitons, {\it Summer School in Modern
Mathematical Physics}, Institute of Physics, Belgrade, SFIN {\bf
A3} (2002) 283-296.
\bibitem{dragovich16} B. Dragovich, On $p$-Adic and Adelic
Generalization of Quantum Field Theory, {\it Nucl. Phys. B (Proc.
Suppl.)} 102, 103 (2001) 150-155.
\bibitem{khrennikov1} A. Khrennikov, {\it $p$-Adic Distributions in
                 Mathematical Physics}, Kluwer AP,
                 Dordrecht, 1994.
\bibitem{khrennikov2} A. Khrennikov, {\it Non-Archimedean Analysis: Quantum
         Paradoxes, Dynamical Systems and Biological Models}, Kluwer AP,
         Dordrecht, 1997.




\end{thebibliography}
\end{document}